\pdfoutput=1

\documentclass[%
reprint,
showkeys,
nofootinbib,
 amsmath,amssymb,
 aps,
prl,
]{revtex4-1}

\usepackage{graphicx}
\usepackage{dcolumn}
\usepackage{bm}

\usepackage{xcolor}

\begin{document}

\preprint{APS/123-QED}

\title{Design of elastic networks with evolutionary optimised long-range communication as mechanical models of allosteric proteins.}

\author{Holger Flechsig}
\email{holgerflechsig@hiroshima-u.ac.jp}
\affiliation{Department of Mathematical and Life Sciences, Graduate School of Science, Hiroshima University, 1-3-1 Kagamiyama,
Higashi-Hiroshima, Hiroshima 739-8526, Japan}

\date{February 24, 2017}

\begin{abstract}
Allosteric effects are often underlying the activity of proteins and elucidating generic design aspects and functional principles which are unique
to allosteric phenomena represents a major challenge. Here an approach which consists in the {\it in silico} design of synthetic structures which,
as the principal element of allostery, encode dynamical long-range coupling among two sites is presented. The structures are represented by elastic
networks, similar to coarse-grained models of real proteins. A strategy of evolutionary optimization was implemented to iteratively improve 
allosteric coupling. In the designed structures allosteric interactions were analyzed in terms of strain propagation and simple pathways which 
emerged during evolution were identified as signatures through which long-range communication was established. Moreover, robustness of allosteric
performance with respect to mutations was demonstrated. As it turned out, the designed prototype structures reveal dynamical properties
resembling those found in real allosteric proteins. Hence, they may serve as toy models of complex allosteric systems, such as proteins.
\end{abstract}


\maketitle

\section*{Introduction}
The functional activity of proteins and its precise regulation often relies on allosteric coupling between different functional
regions within the macromolecular structure. According to the mechanical perspective allosteric communication originates
from structural changes mediated by a network of physically interacting residues \cite{tsai_14}. Much resembling the occurrence of a
{\it proteinquake}, local conformational motions initiated, e.g., upon ligand binding to one specific site propagate across the
protein structure to spatially remote regions eventually generating a functional change in the conformation of another site.

Experiments have indeed evidenced the existence of communication networks in proteins, which are formed by only a set
of amino acids and constitute allosteric pathways, physically linking remote binding sites \cite{bruschweiler_09,grutsch_16}.
Various computational strategies aimed at predicting such pathways have been developed, including structure-based network
analysis \cite{delsol_06,delsol_07,daily_08}, stochastic Markov modelling \cite{chennubhotla_06,chennubhotla_07}, and
sequence-based statistical methods \cite{lockless_99,suel_03,dima_06,tang_07}. Understanding of allosteric communication
at the molecular level has also been widely addressed at atomistic resolution in molecular dynamics (MD) simulations
\cite{ma_00,ghosh_07,cui_08,dixit_11,laine_12, naithani_15,hertig_16}.
 
To circumvent the heavy computational burden present in MD simulations, collective conformational motions and allosteric
transitions in proteins have been, to a vast extend, investigated at the coarse-grained level using elastic-network models and
the analysis of normal modes \cite{xu_03,zheng_06,chennubhotla_08,yang_09,morra_09,erman_13,krieger_15}. Despite their
approximate nature such simplified descriptions have significantly contributed in understanding the mechanistic underpinnings
of allostery in proteins \cite{bahar_07,bahar_10}.

While in the majority of cases the important question is considered to be {\it how allostery works} for a particular protein given
its specific structure, there may also be more fundamental and general questions to be addressed such as what are generic
design and functional principles, requisite to {\it make allostery work}, and, which dynamical properties are unique to allostery.
One possibility to approach such aspects would consist in a systematic screening of the biophysical properties among allosteric
proteins with available structural data, which has already been sporadically attempted earlier \cite{daily_07,daily_08}.

Here, in an attempt to address this subject, we present an alternative approach which is motivated by the idea to engineer, {\it in silico}, artificial spatial structures
with dynamical properties resembling those found in real proteins. In previous works such an approach was employed to design structures of elastic
networks which can operate as a model molecular machine \cite{togashi_07,cressman_08,huang_13} or swimmer \cite{sakaue_10}.
Moreover, we recently used this machine to construct a model motor which in its function mimics the myosin protein
motor \cite{sarkar_16}.

In this paper we aim to establish a generalised structural model of an allosteric system. To this end we design through
evolutionary optimisation elastic-network structures which, as the principal element of allostery, encode long-range coupling among
two spatially remote local regions. We first explain how the strategy of iterative evolution was developed and applied to stepwise
improve, starting from a random elastic network, the allosteric response in the emerging structures. In the designed
structures allosteric communication was then analysed in terms of the propagation of strain and its spatial distribution was
used to identify pathways through which remote interactions are established. Moreover, the effect of mutations was
demonstrated and robustness of allosteric performance in the designed structures examined. Finally we discuss the
relevance of our model system in the light of actual allosteric proteins.

\section*{Results}

\subsection*{Random elastic network and Evolutionary optimisation}

Our structural model of an allosteric system was based on evolutionary optimisation starting from a random elastic network.
The initial random network was constructed as a two-domain structure. It was obtained by randomly folding two polymer
chains independently, each consisting of $100$ elastically linked identical beads and representing one domain, and then
merging them such that they form a common domain interface. In the model physical interactions between all beads,
beyond those acting between neighbouring beads in each chain, were introduced by connecting those two beads by an
elastic link which have a separation smaller than a prescribed interaction radius. Thus the complete elastic network was
obtained. Details of the construction are described in the Methods section.

When elastic networks are constructed based on the structures of real proteins, the beads would corresponds to atoms or
typically represent entire amino acid residues, and elastic links between them empirically mimic effective interactions between
them \cite{tirion_96,haliloglu_97}.  

Here, we proceed with the constructed random network of two compactly folded coupled domains shown in Fig. 1. In each network
domain a pocket site, for simplicity formed by two beads only, was chosen such that a sufficient separation of the two sites was
ensured (see Methods). One site represented the allosteric pocket whereas the other mimicked the regulated pocket, respectively.
As shown in Fig. 1, the pockets were located opposite to each other near the surface of the respective domain.

\begin{figure}[t!]
\centering{\includegraphics[width=8cm]{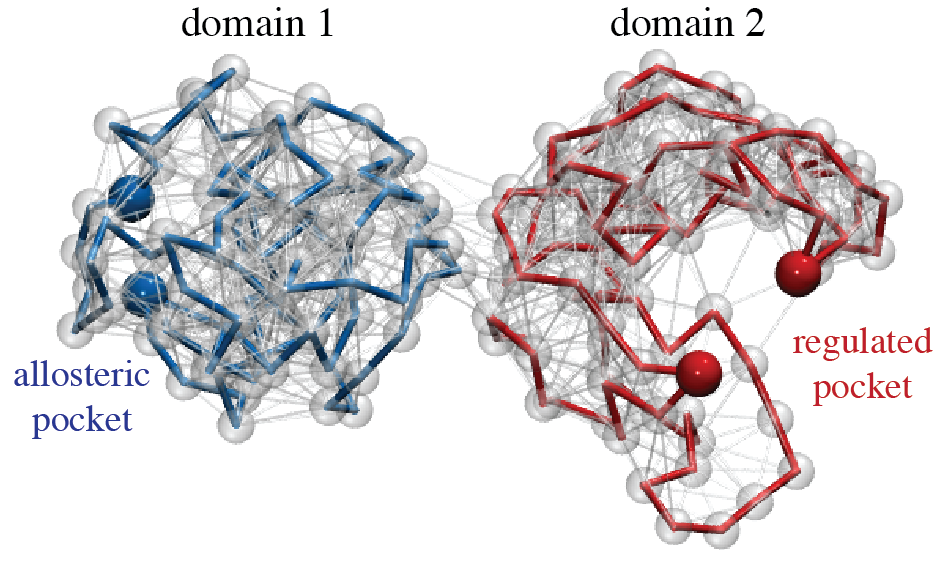}}
\caption{{\bf Random elastic-network structure.} The initial two-domain network consisting of $200$ beads connected by elastic
springs. The randomly folded chains, each forming one domain, are shown as blue, respectively red, traces of thick bonds.
Other springs are displayed as thin bonds in grey. In each domain the two beads representing the remotely placed pockets
are highlighted as coloured beads.}
\end{figure}

With this setup, the ability of the elastic network to conduct allosteric communication between the pockets was examined
using a simple force-probe scheme, in which conformational dynamics in the allosteric site was initiated through local binding
of an additional ligand bead, and the subsequent response generated in the regulated site has been detected. In particular,
attractive pair forces which were acting on the pocket beads of the allosteric site have been applied to mimic binding of the
fictitious substrate bead to its centre. The dynamics of protein elastic-networks consists in processes of over-damped
relaxation motions (see Methods). Hence, as a result of such forces, all beads underwent coupled relaxation motions
bringing the elastic network from its original conformation (without the forces) to a deformed steady state of the network,
having the ligand bead tightly bound to the allosteric site (i.e, with the forces applied for a long time). During this process
the additional forces tend to close this site, i.e. its two beads move towards each other, and first the elastic links localised in
their vicinity became deformed. Eventually deformations propagated through the entire network structure until a final steady
network conformation was reached. In the simulations the conformational motions inside the network were followed by
numerically integrating the equations of motion for all network beads (see Methods). The response generated inside the
regulated pocket in the second domain was quantified in terms of distance changes between its two corresponding beads.

\begin{figure*}[t!]
\centering{\includegraphics[width=12cm]{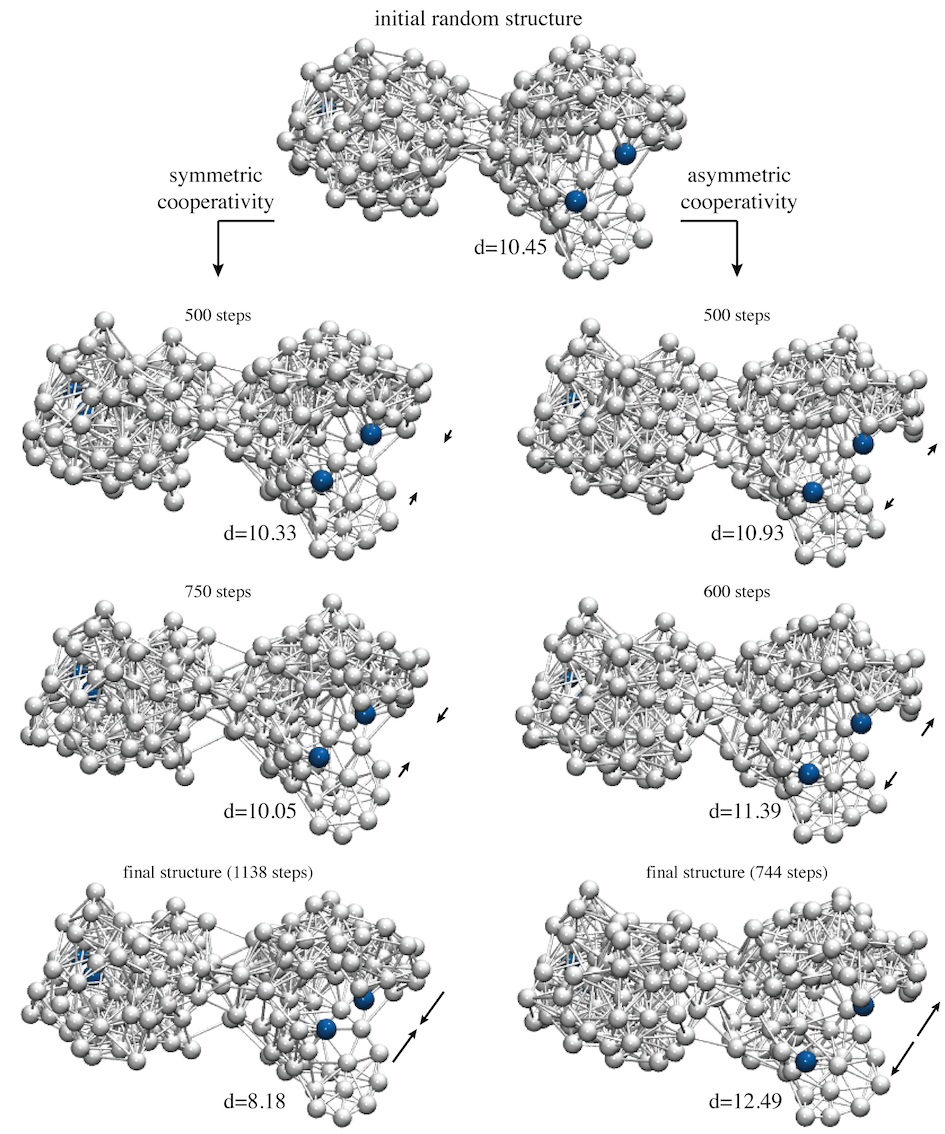}}
\caption{{\bf Evolutionary optimisation.} Snapshots of network conformations taken at different steps of accepted mutations
along the processes of evolutionary optimisation towards designing the two prototype allosteric networks. The initial random
network structure is shown in the top row. Each depicted network conformation corresponds to the steady state with the allosteric
pocket being closed after ligand binding and the allosteric response in the regulated pocket becoming progressively improved.
In the design process of the network with symmetric cooperativity the regulated pocket increasingly closed as evolution proceeds,
whereas during the design of the network with asymmetric cooperativity its ability to open emerged (indicated by black arrows).
Pocket beads are highlighted in blue colour and the size of the regulated pocket in \AA\ is given for all snapshots. Mutation-induced
structural changes and concomitant remodelling of elastic interactions can be seen by comparing snapshots. Final structures of
designed networks are shown in the bottom row.}
\end{figure*}

As we found, the random network structure did not reveal any allosteric coupling between the two sites. While the allosteric
site closed upon substrate binding, with the distance between pocket beads changing by approx. $-4$\AA, no response was
detected at the regulated site with a change of $<10^{-4}$\AA\ (for the choice of lengths units, see Methods).
Proceeding with this observation, an algorithm of evolutionary
optimisation aimed to design networks with pronounced allosteric communication was established. The optimisation process
consisted in sequences of mutations followed by selection, which were applied iteratively starting from the initial random network.
In each evolution step the following sequence was carried out: {\it i) mutation;} a single structural mutation was performed by
randomly selecting one network bead (excluding one of the four pocket beads) and changing its equilibrium position, which
corresponded to an alteration of elastic connections in the vicinity of the mutation spot. {\it ii) probing of allostery;} the
force-probe scheme was applied in a simulation of the new mutant network structure and its conformation with the ligand
bead bound to the allosteric site was obtained. {\it iii) selection;} the allosteric response in the regulated pocket of the network
before and after the mutation was compared and the mutation was scored favourable and was accepted if the mutant network
showed improved allosteric coupling between the two sites; otherwise it was rejected. The optimisation procedure was
iteratively applied and terminated, once a network structure with a prescribed level of sufficient allosteric coupling was obtained.
Details of the implementation are described in the Methods section.

\begin{figure*}[t!]
\centering{\includegraphics[width=16cm]{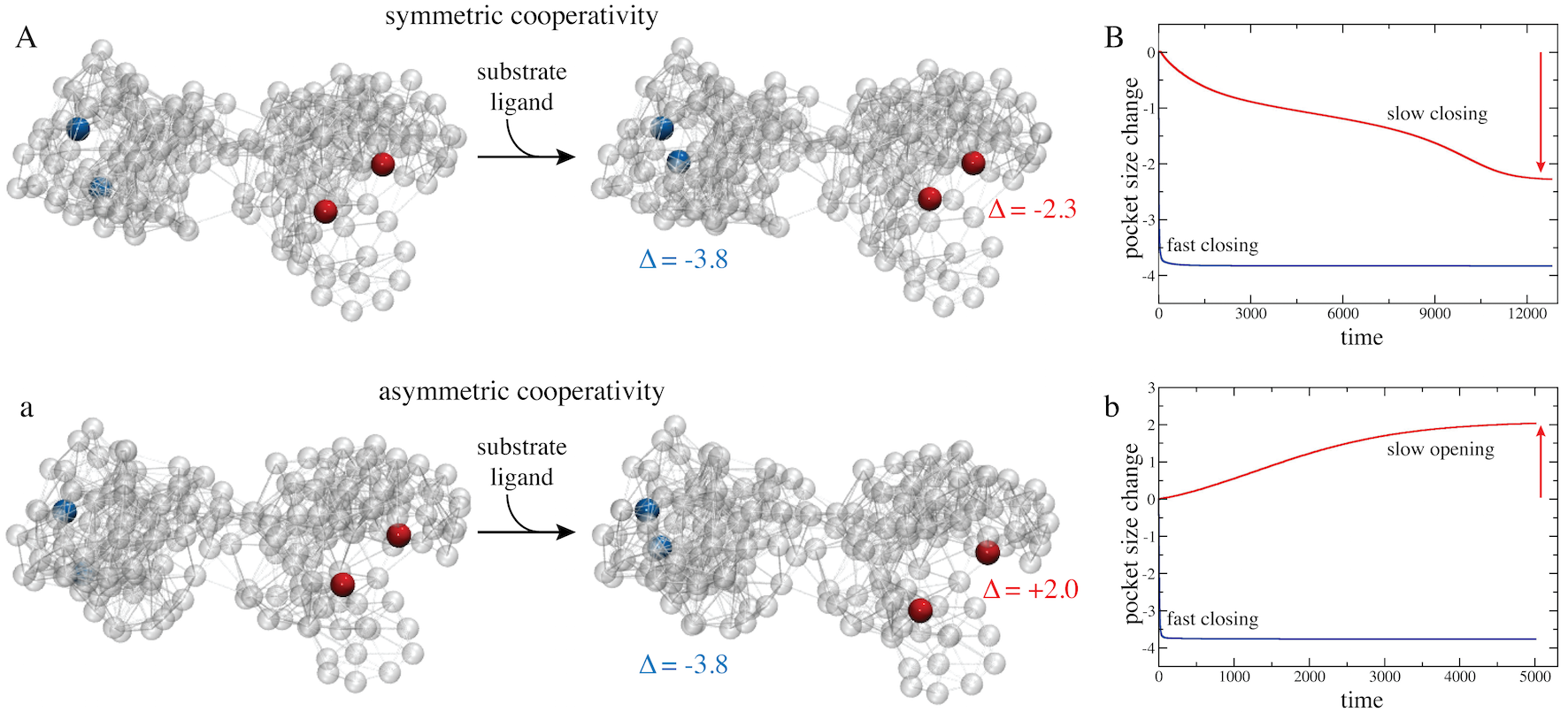}}
\caption{{\bf Allosteric coupling in designed network structures.} (A,a) For both designed networks the initial structure and the
final structure with the substrate ligand bound to the allosteric pocket are shown. Elastic connections between network
beads are shown as thin transparent lines and pocket beads are coloured blue (allosteric pocket) and red (regulated pocket),
respectively. In both designed networks the initial and respective final structure is very similar (see text), but the large-amplitude
motions of pocket beads evidencing intra-structure allosteric communication are apparent. In the network with symmetric
cooperativity substrate-induced closing of the allosteric pocket triggers allosteric closing of the regulated pocket, whereas
in the network with asymmetric cooperativity the allosteric response consisted in opening of the regulated site. (B,b) Traces
showing changes in the size of both pockets upon binding of the substrate bead to the allosteric pockets are also provided
(blue colour for allosteric the pocket, red for the regulated pocket).
}
\end{figure*}

In this work two prototype examples of elastic network structures with allosteric coupling were designed and analysed. 
In two independent simulations of evolutionary optimisation, starting both times with the same initial random network, two
different network structures with optimised allostery have been designed. The first one was designed under the side condition
that allosteric coupling between the remote sites had the type of {\it symmetric cooperativity}, i.e., binding of the substrate
bead to the allosteric site and the concomitant closing of that pocket resulted in consequent closing of the regulated site.
In the second case, the requirement was vice versa; allosteric communication was optimised under the premise that
substrate induced closing of the allosteric pocket triggers opening in the regulated site, i.e. {\it asymmetric cooperative}
coupling was realised. In both cases, as a result of several hundreds of mutations which were needed during evolution
to improve allosteric coupling, the networks underwent significant structural remodelling and concomitant rewiring of
elastic interactions between their beads. In Fig. 2 snapshots of network architectures along the processes of evolution
are displayed. For both designed networks it could be observed that besides mutation-induced changes taking place
inside the individual domains, in particular their common interface became significantly remodelled. As evolution
progressed the allosteric response generated inside the regulated pocket upon ligand-binding to the allosteric
pocket was gradually enhanced during both design processes, as can be seen by comparing corresponding snapshots
of network conformations in Fig. 2. In the next paragraphs the two prototype designed network structures are presented
and their dynamical properties analysed.

\subsection*{Designed prototype networks}

The designed networks are shown in Fig. 2 (bottom panel). Both networks retained the initial two-domain architecture after
the evolution, their structures, however, were clearly different from the initial random network. In both designed networks a
pronounced domain interface with many inter-domain links had emerged; such interface was rather sparse in the initial
random network (see Fig. 2). Remodelling of that region under the process of evolution apparently points towards the
importance of the pattern of elastic connections in the domain contact region for allosteric communication in our model
systems. Those aspects will be further discussed in the next section.

In Fig. 3 (A,a) the two designed networks are shown each in their equilibrium conformation and in the respective steady conformations
with the ligand bound to the allosteric pocket and the allosteric response having been triggered in the remote regulated
sites. Traces showing changes in the size of both pockets upon binding of the ligand bead to the allosteric pockets are also
provided (see Fig. 3B,b). In both designed networks the allosteric pocket closed rapidly when the substrate bead bound there, with
the size changing by $-3.8$\AA\ in both cases. In striking contrast, motions inside the regulated pocket were much slower, when
changes in its size only gradually set in and the pocket smoothly approached its closed conformation in the network with
{\it symmetric cooperativity} (change by $-2.3$\AA), or open conformation in the network with  {\it asymmetric cooperativity},
respectively (change by $+2.0$\AA). Movies visualising conformational motions during the allosteric transition in the designed
networks are provided as Supplementary Videos V2 and V3.

\begin{figure*}[t!]
\centering{\includegraphics[width=17cm]{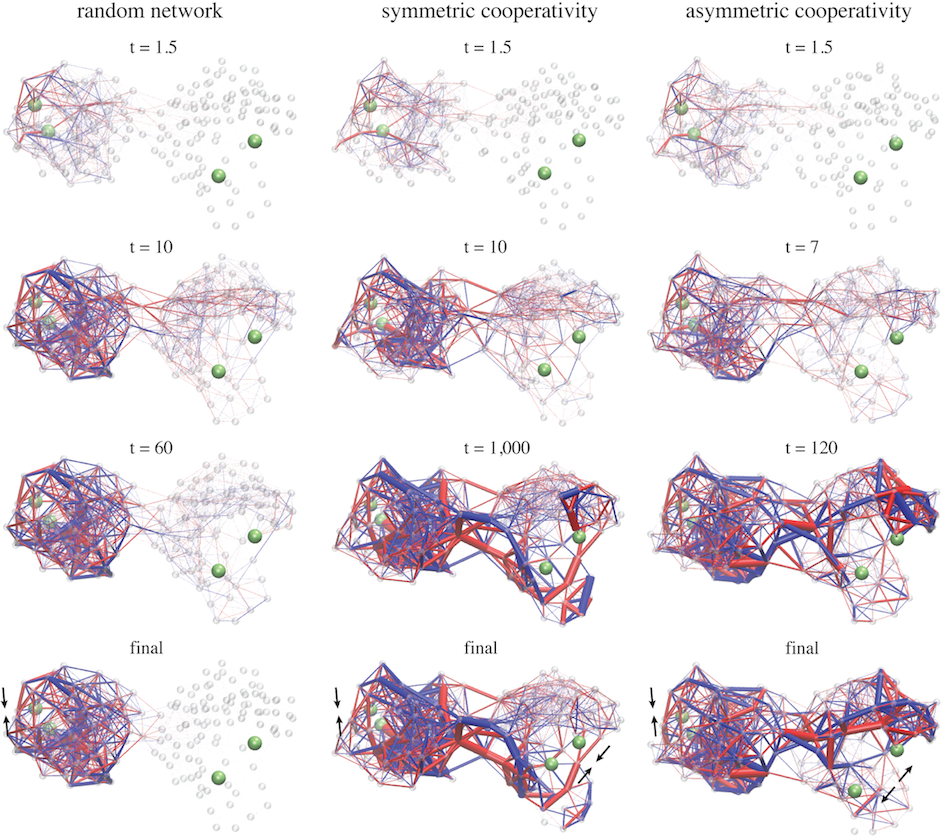}}
\caption{{\bf Strain propagation.} Propagation of strain subsequent to ligand-binding to the allosteric pocket is visualised in the
random network and in the two designed structures. Snapshot taken at different time moments during each simulation are shown
in similar perspectives. Network bonds are displayed bicolored, blue indicating stretching of the bond and red denoting compression
as compared to the corresponding natural bond length. The bond thickness encodes the absolute value of strain the bond was subject to.}
\end{figure*}

Comparing the ligand-bound network conformation with the initial structure reveals that in both designed networks the
conformational changes underlying allosteric dynamics did not involve any large-scale structural rearrangements but was rather
governed by small-amplitude subtle motions. In fact, initial and substrate-bound structures compared by RMSD's of only $0.6$\AA\
in the symmetric case and $0.46$\AA\ in the asymmetric case, respectively. While local motions in the two pockets were indeed
pronounced, the communication between them must therefore have been resulted from cascades of small-amplitude
displacements of network beads located in the region connecting both pockets.

By the application of evolutionary optimisation to the random network we could successfully design special networks whose
two-domain structure encodes enhanced allosteric interactions between two remote pockets. All three network structures are
provided as Supporting Data.
Next, to unravel signatures which may underly allosteric communication in the two prototype networks we have analysed the
conformational dynamics in terms of mechanical strain propagating through the network structures.

\subsection*{Propagation of strain and communication pathways}

Apparently, allosteric interactions in our model systems result from conformational changes propagating from the allosteric
site across the domain interface to the regulated site. Therefore, in both designed structures the elastic strain inside the networks
was computed during the entire simulation following ligand binding to the allosteric site until the steady state of the respective network
was reached. The strain of a network was determined in terms of the deformation of elastic links connecting the beads, i.e. deviations
from their natural lengths in the initial ligand-free network. Details of the computation are found in the Methods section and SI text.
To conveniently visualise strain propagation, the network links are displayed bicoloured, to distinguish stretching or compression,
and their width is proportional to the magnitude of the respective deformation. In order to compare to the designed networks the strain
propagation was also followed in the initial random network which was allosterically inactive. Snapshots of the strain distribution in all
three networks taken at different time moments during the simulation are shown in Fig. 4. Movies visualising conformational motions
and the propagation of strain are provided as Supplementary Movies V1-V3.

\begin{figure}[t!]
\centering{\includegraphics[width=8cm]{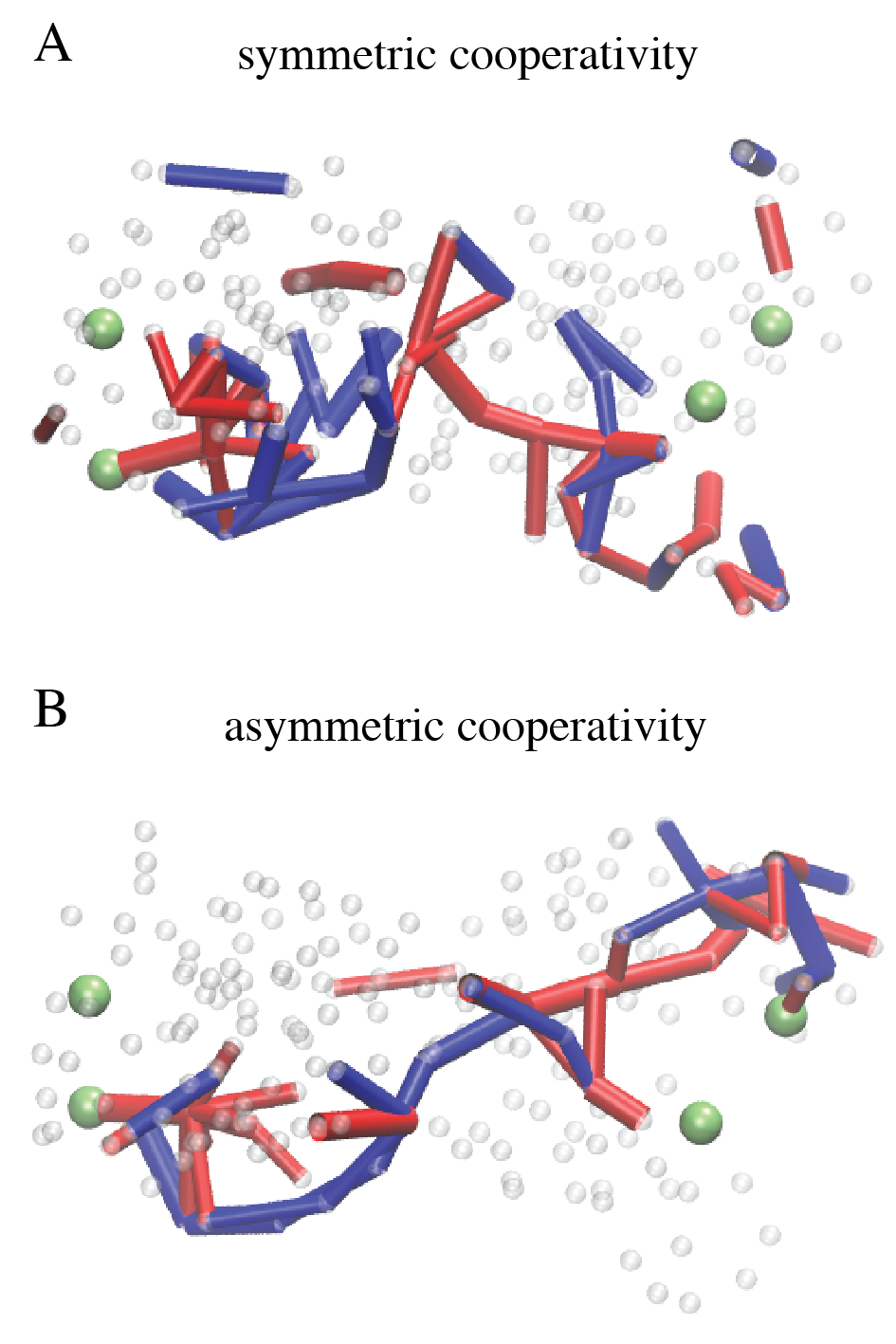}}
\caption{{\bf Communication chains in designed structures.} Elastic bonds which are significantly involved in the transport of strain
are found to form simple connected chains which meander through the structure and connect the allosteric pocket in domain $1$ (green beads, left side)
with the regulated pocket in the remote second domain (green beads, right side).}
\end{figure}

In all three networks the strain was first localised in the vicinity of the allosteric pocket where the ligand was bound. After that,
rapidly interactions between other beads also set in and the majority of links in domain $1$ became deformed. In the random
network the distribution of strain in domain $1$ was rather homogeneous; most links were stretched or compressed at similar levels
and there was no distinctive feature present. In this network any significant deformations of links located in the central domain interface
region could be detected and therefore spreading of strain into the second domain during the simulation was practically absent.
The structure of this domain is apparently very stiff preventing any internal conformational motions to occur. Hence, the random
elastic network cannot conduct allosteric communication. The situation in both designed networks was completely different.
There, first, after ligand binding to the allosteric site conformational motions spread across the domain interface and
generated deformations in the second domain and in the vicinity of the regulated site. Second, not all elastic pair interactions
in the designed structures were equally involved in the process of allosteric coupling. Rather, the temporal distribution of strain
shows that communication between the remote pockets proceeded through specific sub-networks which were apparently critical
for allostery in the designed structures. Those networks are formed by a subset of beads connected by springs which undergo
major deformations and, hence, contribute essential pathways for the spreading of conformational motions from the allosteric
site across the structure to the regulated site. In both designed network structures such remarkably strained springs were found
in domain $1$ and, in contrast to the random network, in the central domain interface region and the second domain, where they
form sparse clusters of important pair interactions between beads.

\begin{figure*}[t!]
\centering{\includegraphics[width=17cm]{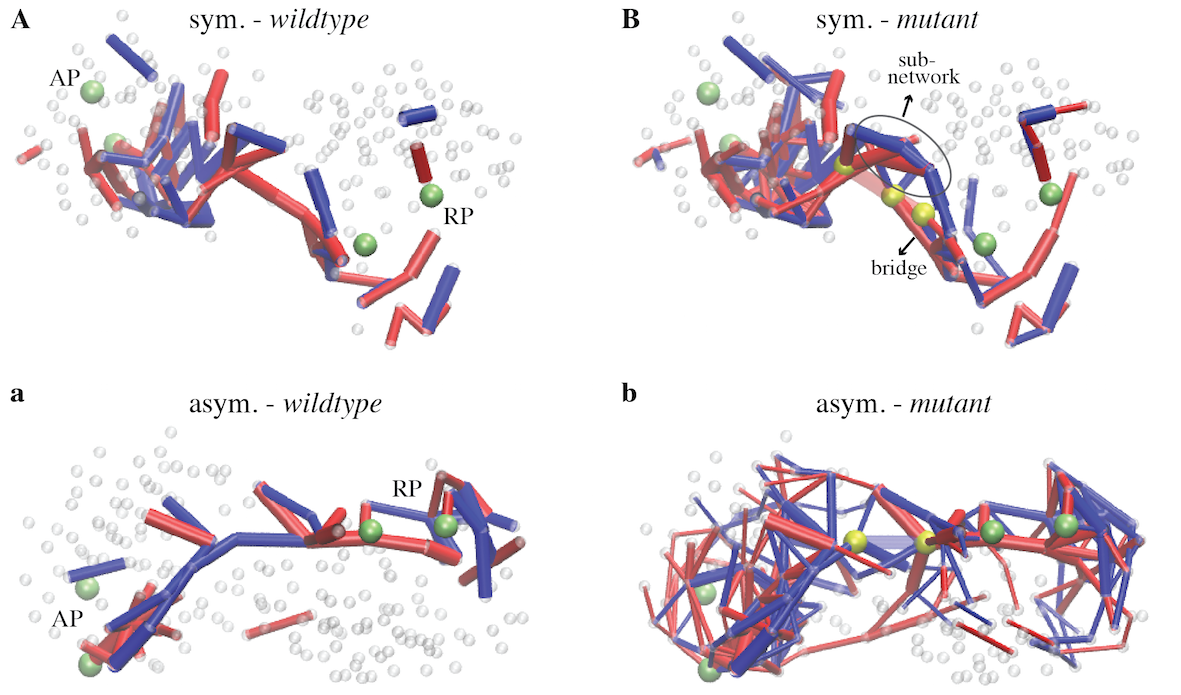}}
\caption{{\bf Communication pathways in {\it mutant} structures.} Communication chains in the designed {\it wildtype} network structures
(A,a) are compared with the communication pattern obtained from strain propagation in a corresponding selected {\it mutant} structure (B,b).
Communication chains in (A,a) are the same as shown in Fig. 5 but here another perspective is chosen for better comparison with the
{\it mutant} structures. B) Communication pathways in the $52-197-196$ double mutant of the network with symmetric allosteric coupling.
Beads with index $52$,$197$ and $196$ are highlighted in yellow and the two bonds removed between them are shown in transparent. 
Links which in this mutant became important to redirect strain over the mutation site are indicated (by an ellipse). b) Communication pathways in the $52-185$
mutant of the designed network with asymmetric allosteric coupling. Beads $52$ and $185$ are displayed in yellow and the deleted link between
them is transparent. Allosteric communication in this mutant was maintained via a complex strain transportation network and a simple chain
as found in the {\it wildtype} was not available (compare a,b).
}
\end{figure*}

Regions in which the elastic springs remained only marginally populated by strain were rather stiff as compared to the rest of
the network. In that regard we observed that the structure of domain $2$ in the designed networks is special. In the network with
symmetric allosteric cooperativity the two lobes in the tweezer-like domain structure, each harbouring one of the pocket beads
of the regulated site, were found to be quite flexible whereas the other parts in domain $2$ were stiff (see Fig. 4). In the network with asymmetric
allosteric coupling the hinge region connecting both lobes appears to be very flexible instead (see Fig. 4). The specific pattern of flexibility
in the designed networks has apparently emerged during the process of evolutionary optimisation in order to make opening/closing
motions of the regulated site possible and hence enable the designed network architectures to conduct allosteric communication.

The analysis of strain propagation in the designed networks clearly reveals functionally important signatures underlying the
investigated types of allosteric coupling. Nonetheless the strain distribution was rather complex and we therefore aimed to deduce
simpler pathways which could characterise communication between the pockets, by focusing only on those network springs
which carry major strain (see Methods). Results are shown in Fig. 5. In both networks simple chains formed by significantly
deformed adjacent network springs were found to meander from the allosteric pocket in domain $1$ via the domain interface
to the regulated pocket in domain $2$. Such chains are composed of a series of stretched springs in domain $1$, potentially
caused by the closing motion of the allosteric site, connected to a linkage of compressed springs located in domain $2$ and
triggering the respective allosteric response there.

The identified communication chains are obviously critical for the transmission of conformational changes between the two
domains and therefore may represent a functional skeleton, critical for allosteric activity in the designed network architectures.
Immediately one may pose the question whether, and to what extend, allosteric coupling can be maintained if structural changes
would be applied to such motifs.

\subsection*{Robustness of allosteric communication}

With regard to the robustness of allosteric communication in the two designed networks the intention was not to perform
a systematic screening of the effect of structural perturbations. Instead, the aim was rather to present a demonstration. Therefore
the analysis of robustness was limited to a few exemplary structural mutations, those applied to beads belonging to the identified
communication pathways, and on examining their influence on allosteric interactions.
In a first set of simulations mutations which consisted in the deletion of only a single elastic link between two network beads were
considered. In the realm of proteins such kind of perturbation would roughly correspond to the mutation of a single amino acid residue,
resulting in specific local interactions to disappear. For each specific {\it mutant} network a single simulation starting with ligand binding to the allosteric pocket
was carried out and after completion the response in the regulated pocket was quantified and the level of allosteric communication determined.
The results are listed in Supplementary Table T1. It is found that allostery in both designed networks is generally robust with respect to the removal of single
interactions. To understand how long-range coupling between both pockets is maintained in the {\it mutant} networks we have visualised the propagation
of strain and determined communication pathways for some exemplary cases, similar to what has been performed for the designed {\it wildtype} networks
(see previous section).

For the network with symmetric cooperativity three cases were focused on in more detail, all of which maintained allosteric activity (see Supplementary Table T1). They
corresponded to a mutant with one major interaction at the central domain interface having been removed and another mutant with a link
deleted in domain $2$, closer to the regulated pocket. The computed communication chains in those two mutants are shown in Supplementary
Fig. S3. Movies of strain propagation in those two mutant networks are provided as Supplementary Movies V4 and V5.
To highlight robustness in the first designed network, a third example consisting in a double mutant network in which both selected links were deleted, was
considered too (see Supplementary Movie V6). For all three mutant networks we find that the strain propagation networks as well as the
extracted communication chains are very similar to that of the wildtype networks, except in the vicinity of the respective mutation site (see Fig. 6A,B and Fig. S3).
There it is found that in the neighbourhood of the deleted link a few bead interactions, which in the wildtype network have not played a major
role in the transmission of strain, became important, forming a bridge through which strain was able to side-track the mutation site (see Figs. 6B and
S3).
This result demonstrates that in the mutant structures similar communication pathways are available which, together with the bridge motifs
that compensate the mutation defect, are employed for the propagation of conformational changes and hence may provide the foundation
for the robustness of allosteric coupling in this designed network.

For the designed network with asymmetric allosteric coupling one mutant structure with a single deleted link located at the domain interface
has been selected for illustration. Despite the mutation it revealed full allosteric activity. Propagation of strain subsequent to ligand binding to the
allosteric pocket is shown in Supplementary Movie V7 and the extracted communication pathways are depicted in Fig. 6b. It is found that 
the network of bead interactions through which communication between both pockets was transmitted is rather complex, with the strain being
populated in large parts of the {\it mutant} structure. Hence, in contrast to the {\it wildtype} network, allosteric coupling in this mutant does not proceed
via simple communication chains.

While generally allosteric communication in both designed networks was found to be robust with respect to the deletion of a single bead interaction,
there were still few critical mutations which involved a drastic decrease in the coupling between both pockets. They can correspond to perturbations
in the vicinity of either pocket or located at the domain interface (see Supporting Figs. S2A,B). In the final step of investigations the effect of stronger
mutations was considered, where an entire bead was deleted in a simulation of each network. Those perturbations typically led to a remarkable
drop or even complete knockout of allosteric communication (see Supporting Table T1).

\subsection*{Summary and Discussion}

The phenomenon of allosteric coupling between different functional regions within a macromolecular structure is ubiquitously present
in proteins and therefore raises important questions of the fundamental nature of the underlying mechanisms. In contrast to typically
employed structure-based modelling of protein dynamics, a different methodology towards approaching such essential problems is
presented in this paper. Instead of considering real protein structures, artificial analog structures which encode pronounced long-range allosteric
coupling between two spatially remote pockets were designed. The structures were represented by elastic networks, similar to
coarse-grained models widely used to describe conformational dynamics of real proteins.

A force-probe scheme consisting in ligand-binding to the allosteric pocket, following conformational motions spreading over the
structure, and detecting the response generated in the regulated pocket was implemented to evaluate the ability of the network
to conduct allosteric communication. Initially starting with a random elastic-network, which did not reveal any allostery, a scheme
of evolutionary optimisation was iteratively applied to design two prototype elastic-network structures with {\it perfected} allosteric
coupling, one with symmetric and the second with asymmetric cooperativity.

In the designed networks well-defined pathways and simple chains of important interactions, established by only a few network
beads, were identified to constitute the signatures which underly allosteric communication. Remote interactions were robust even
if minor structural perturbations were applied. However, a single critical mutation could knockout completely allosteric communication
in the designed networks.

While the first descriptions of allosteric systems, namely the MWC and the KNF models \cite{monod_65,koshland_66}, were formulated
more than 50 years ago in the absence of any structural data and were thus of phenomenological nature, the development of very
sophisticated experimental techniques and the vast amount of protein structures becoming available in the last decades have ever
since allowed to investigate allosteric communication in proteins on the molecular level and led to a decent understanding of the
mechanism underlying allostery in several role model proteins. In the recent past structure-based computational modelling,
aiming to follow the conformational motions as the underpinning of allosteric effects in proteins, has gained a lot of attention. In
particular, due to the time-scale gap present in atomistic-level molecular dynamics simulations, coarse-grained elastic network
models which are limited to the mechanical aspect of protein operation became very popular to investigate slow allosteric transitions
with timescales beyond the microsecond range. In those studies the analysis of conformational changes is typically based on the
computation of normal modes and allosteric effects are discussed in terms of short- and long-ranged correlations of amino-acid
residue displacements.

The aim in the present study was to present a model which emphasises the mechanical picture of allosteric systems. Therefore,
also the elastic network model was used. Here, however, the full elastic dynamics of the network was considered by always
numerically integrating the non-linear equations of motion and monitoring processes of conformational relaxation; no linearisation
was performed and the conformational dynamics beyond the limiting normal mode approximation could be followed.  As a
consequence, this model has the obvious advantage that it can resolve the temporal order of events which eventually establish
allosteric communication in the network structures, starting from forces and strains which were first generated locally at the allosteric
site as a consequence of ligand binding, followed by the subsequent propagation of conformational motions via the domain interface,
to finally induce structural changes in the remote regulated pocket. The current model therefore naturally includes causality as the
guiding principle to manifest allosteric communication and therefore is richer in its explanatory power compared to correlation-based
analyses.

The presented model emphasises the structural viewpoint of allostery (as described in \cite{tsai_14}) in which allostery is regarded
as a consequence of optimised communication between the remote functional sites, established through the propagation of strain
along pathways which are formed by a set of interacting residues. The employed elastic network description clearly implies limitations.
All network particles are of the same kind and physical interactions between them are incorporated by empirical effective potentials;
thermal fluctuations were also neglected for simplicity.
Despite such gross simplifications the dynamical properties of the designed prototype allosteric structures reveal remarkable
similarities with those found in real allosteric proteins: i) specific parts of the structure are flexible whereas other regions form
stiff clusters; ii) the propagation of conformational changes which results in the long-range coupling of the remote sites proceeds through
communication pathways involving only a few of the many intra-structural interactions; iii) single critical mutations can knockout
allosteric coupling. In summary, the designed elastic network structures can provide a general physical model for the mechanics of complex allosteric
biomolecules, such as proteins.

It should be remarked that the evolutionary pressure imposed in the design process consisted only in magnifying changes in the
regulatory pocket in response to ligand-induced changes in the allosteric pocket; no other requirements were imposed and the
dynamical properties which actually improved allosteric communication in the evolving structures emerged autonomously.
However, in future studies design algorithms which involve coevolution can also be implemented and multiple optimisation
criteria can be imposed. Moreover, in future studies the design and analysis of a larger number of allosteric elastic networks can be
undertaken. That would allow to compare properties such as communication pathways among them and possibly draw conclusions
on the generality of such dynamical motifs.

Previously the relaxational elastic-network approach employed in this study was already applied to investigate allosteric coupling in
helicase motor proteins \cite{flechsig_10,flechsig_11} and in the myosin-V molecular motor \cite{duttmann_12}. In those studies,
however, allosteric coupling was only qualitatively discussed. The analysis of strain propagation performed in this paper for the
designed artificial protein structures, and the methods to quantify and visualise communication pathways, can easily be applied in
the structure-based modelling of real allosteric proteins and allow to investigate intramolecular communication in dynamical
simulations.

{\bf Added note} After finishing the manuscript I became aware of two quite recently appeared works in which also the design of mechanical
networks with allosteric coupling was undertaken (published in PNAS \cite{yan_17,rocks_17}). There, however, different computational
algorithms were used and optimisation of allosteric coupling was performed within the linear response limitation; the underlying
network architectures were also very different (e.g. 2D on-lattice models \cite{yan_17}). Both works principally demonstrate
the applicability of design processes to obtain desired responses. 

The design process introduced in my manuscript was developed completely independent. Besides other differences, its principal
distinction is that during the design process the allosteric response was optimised by always considering the full nonlinear elastic
dynamics of the networks. The importance of nonlinearities for protein function involving allostery has previously been evidenced
(e.g. in \cite{miyashita_03,togashi_10}). Secondly, the network architectures used in the present study resemble more closely the
three-dimensional compact fold of real proteins; in fact the designed networks can be regarded as coarse-grained representations
of {\it fictitious} protein structures. Most importantly this work goes beyond developing the design process. In fact, the main emphasis
here was put on understanding the mechanistic principles underlying long-range communication in the designed networks, which
was achieved by visualising and analysing propagation of strain, extracting communication pathways and chains which were critical for allosteric
coupling, and, investigating robustness of the designed functional properties.

\footnotesize
\section*{Methods}

\subsection*{Construction of random elastic network}
The initial two-domain elastic network was constructed by first randomly folding two chains of linked beads, then
bringing them into contact and finally determine the network connectivity.
One chain consisted of $100$ identical beads each and its folded form was constructed as follows. After fixing the
position of the first bead, each next bead was placed at random around the position of the previous bead, with the
following restrictions: i) the distance to the preceding bead had to lie within the interval between $l_{min}$ and $l_{max}$,
ii) the new bead had to be separated from each previous bead by at least the distance $l_{min}$, and, iii) the distance
from the new bead to the geometric centre of all previous beads should not exceed the threshold $r_{max}$. In the
simulations prescribed values $l_{min}=4.0$\AA, $l_{max}=5.0$\AA, and $r_{max}=20.0$\AA\ were used in order
to generate a compactly folded backbone chain. After constructing two such chains they were merged in such a way
that they came into tight contact but still did not overlap. Positions of the 200 beads in the initial two-domain random
structure are denoted by $\vec{R}_{i}^{(0)}$ and their spatial coordinates are provided as Supporting Data. To
complete the network construction we have checked all distances $d_{ij}^{(0)}=|\vec{R}_{i}^{(0)}-\vec{R}_{j}^{(0)}|$
between beads $i$ and $j$ and introduced an elastic spring between those pairs of beads for which the distance $d_{ij}^{(0)}$
was below a prescribed interaction radius of $r_{int}=9$\AA. With this choice the initial two-domain elastic network had
$1467$ links.
It should be noted that in this model all length units are in principle arbitrary. When data from real protein structures are used,
the distances between amino acids have the scale of Angstroms. Hence, throughout the paper this notion is adopted.

\subsection*{Elastic conformational dynamics}
The total elastic energy of the network is the sum of contributions of all elastic links, i.e. $U=\sum_{i<j}^{N}\kappa\frac{A_{ij}}{2}(d_{ij}-d_{ij}^{(0)})^{2}$.
Here, $N=200$ is the number of beads, $\kappa$ is the spring stiffness constant (equal for all springs), $d_{ij}=|\vec{R}_{i}-\vec{R}_{j}|$ is the actual
length of a spring connecting beads $i$ and $j$ in some deformed network conformation, with $\vec{R}_{i}$ being the actual position vector of bead
$i$, and $d_{ij}^{(0)}$ is the corresponding natural spring length. Coefficients $A_{ij}$ have value $1$ if beads $i$ and $j$
are connected by a spring (i.e. when $d_{ij}^{(0)}<r_{int}$), and equal $0$, otherwise.

The dynamics of the elastic network is governed by Newton's equation of motion in the over-damped limit, where the
velocity of each network bead is proportional to the forces applied to it. The equation for bead $i$ was
\begin{equation}
\begin{aligned}
\frac{d}{dt}\vec{R}_{i}&=-\frac{\partial}{\partial\vec{R}_{i}}U+\vec{f}_{i}\\
&=-\sum_{j}^{N}A_{ij}\frac{d_{ij}-d_{ij}^{(0)}}{d_{ij}}(\vec{R}_{i}-\vec{R}_{j})+\vec{f}_{i}.
\end{aligned}
\end{equation}
On the right side are the elastic forces exerted by network springs which are connected to bead $i$, they only depend on
the change in the distance between two connected beads. Additionally, an external force $\vec{f}_{i}$ could be applied to bead
$i$, which was used in probing allostery (see next section). In the above equations a rescaled time was used to remove
dependencies of the beads' friction coefficient (equal for all beads) and $\kappa$. To obtain the positions of network beads,
and hence the conformation of the network at any time moment, the set of equations of motion was numerically integrated.
In the simulations a first order scheme with a time step of $0.1$ was employed.

\subsection*{Pocket sites and probing of allostery}

In the network model the allosteric site and the regulates site were, for simplicity, each represented
by two beads. Those beads have been selected in such a way that the two pockets were sufficiently remote from each other. At the
same time the two beads forming one pocket should be adjacent, but not directly connected by an elastic spring. In the constructed
two-domain network the allosteric pocket was defined by beads with indices $38$ and $81$, belonging to the first domain, and for the
regulated pocket beads from the second domain, with indices $149$ and $189$, were chosen.

To probe allosteric communication between the two pockets a simple force-probe scheme in which conformational dynamics in
the allosteric site was initiated through the application of additional forces and the subsequent response generated in the regulated
site was probed by evaluating structural changes there. In the simulations pair forces have been applied to the beads of the
allosteric pocket. The forces were always acting along the direction given by the actual positions of the two pocket beads.
Specifically, the force applied to pocket bead $38$ was $\vec{f}_{38}=0.5\cdot\vec{u}$, with the unit vector
$\vec{u}=(\vec{R}_{81}-\vec{R}_{38})/|\vec{R}_{81}-\vec{R}_{38}|$. The same force, but with the different sign, was acting
on the second pocket bead, i.e. $\vec{f}_{81}=-0.5\cdot\vec{f}_{38}$. Such pair forces would induce only internal network deformations
and result in a decrease in the distance between the two pocket beads, thus leading to closing of the allosteric pocket.
Since this situation is apparently equivalent to assuming that an additional network bead becomes bound to the centre of the two
pocket beads, where it generates attractive forces between itself and each pocket bead, we can also say that the chosen force
scheme mimics binding of a fictitious ligand bead to the allosteric pocket.
The action of the additional forces generated deformations of the network which were first localised in the vicinity of the substrate
pocket but gradually spread over the entire network structure. The corresponding process of conformational relaxation was
followed by integrating the equations of motion until a steady state of the elastic network, in which all motions were terminated,
was reached (at final time $T$, see SI text). In the resulting conformation of the network, the effect of allosteric coupling was examined
by evaluating the distance between the beads corresponding to the regulated site. This distance
$d_{149,189}(T)=|\vec{R}_{149}(T)-\vec{R}_{189}(T)|=: A$ is termed the allosteric parameter.

\subsection*{Evolutionary Optimisation}

To design networks with perfected allosteric communication a process of evolutionary optimisation, consisting of mutations followed
by selection, was applied iteratively starting from the random network. In particular the following sequence was carried out.
First the allosteric response of the elastic network before the mutation was determined and the allosteric parameter $A$ stored.
Then a single structural mutation was
performed by randomly selecting one network bead (except for one of the four pocket beads) and changing its equilibrium position. The new equilibrium position was
chosen to be randomly oriented within a sphere of radius $2.0$\AA\ around the old equilibrium position. To preserve the backbone chain
of each domain, it was additionally required that, after the mutation, the distance between the mutated bead and its left and
right neighbour in the chain still lie within the interval between $l_{min}$ and $l_{max}$. Furthermore, distances from the
mutated bead to all other network beads should not be smaller than $l_{min}$. After the mutation the network connectivity $A_{ij}$
was updated by reexamining distances between all beads and the mutated bead; only those pairs which were separated by
a distance less than $r_{int}$ were linked by a spring.
After a mutation the elastic network may posses internal free rotations originating from loosely coupled network parts. They
can lead to local movements free of energetic cost which was to be prevented. Therefore, when the number of non-zero
eigenvalues in the spectrum of the elastic network was smaller than $3N-6$ (indicating the occurrence of internal rotation
zero modes), the mutation was rejected.
Once a mutation which fulfilled all criteria was found, probing of allostery in the new elastic network was proceeded as described
in the previous section, the allosteric parameter after the mutation $A^{mut}$ was determined, and the mutation was evaluated
by comparing the allosteric parameter before the mutation $A$ with that after the mutation $A^{mut}$. Only mutations which were
favourable, i.e. those which improved the allosteric response in the network, were selected. Two situations were distinguished.
In the evolution process where symmetric coupling between the allosteric and regulated pocket was to be optimised, a mutation was
accepted only if $A^{mut}<A$, and rejected otherwise. In the second independent evolutionary process corresponding to the
asymmetric situation, the acceptance criteria for the mutations was $A^{mut}>A$.
As a termination condition for the two evolution processes we imposed $(A-A^{random})<-2.0$\AA\ for the design of the network with symmetric coupling
and $(A-A^{random})>2.0$\AA\ for the design of the network with asymmetric coupling. $A^{random}$ denotes the allosteric parameter
of the initial random network. During both design processes the improvement of allosteric response in the evolving networks
was recorded and it shown in Supplementary Fig. S1.

\subsection*{Strain propagation and pathways}

In the initial random, in the two designed networks, and in the selected mutant networks, the propagation of strain after ligand binding to the allosteric pocket
was monitored. The strain of an elastic link connecting beads $i$ and $j$ was defined as $s_{ij}(t)=d_{ij}(t)-d_{ij}^{(0)}$.
In the employed model conformational changes corresponded to relaxation processes of the elastic network structure (see Equ.'s (1)).
Therefore, the energy injected locally at the allosteric site as a consequence of ligand-binding there would be not only converted into
deformations of elastic bead connections but also dissipate. In particular deformations of elastic springs become significantly damped
the farther away they are located from the pocket.
Since we still wanted to discuss the anisotropy of strain distribution in the network, a method in which the strain was
normalised with respect to the distance from the allosteric site (in terms of the minimal path) was employed. Details
are described in the Supplementary Information. For the visualisation of strain a link was coloured blue
(if $s_{ij}>0$) or red ($s_{ij}<0$) and the width corresponded to $|\tilde{s}_{ij}|$ (superscript $\sim$ refers to the normalised strain).
To determine the communication chains shown in Fig. 5 the maximum absolute strain of each link during the simulation was stored
and only those links whose normalised strain exceeded a prescribed threshold $\tilde{s}_{t}$ were considered (see SI text).
For both designed networks a threshold value of $\tilde{s}_{t}=60\%$ was imposed.
To obtain the communication skeleton in the $52-196-197$ double mutant of the designed network with symmetric
cooperativity (shown in Fig. 6B) a threshold of $\tilde{s}_{t}=45\%$ was used. For the $52-185$ mutant of the designed
network with asymmetric cooperativity (Fig. 6b)  a threshold of $\tilde{s}_{t}=25\%$ was used.

\subsection*{Robustness}

All performed mutations are listed in SI Table T1. For each mutant network a robustness coefficient was computed as the ratio
of the change in the pocket size of the regulated pocket in the considered mutant network and the pocket size change in
the {\it wildtype} network, i.e. $(\tilde{d}_{149,189}^{(0)}-\tilde{d}_{149,189}^{final}) /(d_{149,189}^{(0)}-d_{149,189}^{final})$.
Here, superscript $\sim$ denotes distances in the mutant network and superscript $final$ denotes distances in the corresponding steady
state of the elastic network after ligand binding to the allosteric site (i.e. at time $T$).

\normalsize
\section*{Acknowledgements}
The author is grateful to Alexander S. Mikhailov and Yuichi Togashi for helpful discussions. This work is supported by JSPS
KAKENHI Grant Number 16K05518. All figures of networks and the movies have been prepares using the VMD software \cite{humphrey_96}.

\newpage

\renewcommand{\figurename}{FIG. S}
\setcounter{figure}{0}

\section*{Supplementary Information}

\section*{Probing of allostery and evolutionary optimisation}

The force-probe employed scheme to score the allosteric response in the networks during the design processes was explained in the main text.
To follow conformational motions in a network subsequent to ligand-binding to the allosteric site, the set of equations (Equ.'s (1), main text)
was numerically integrated, always until a steady state of that network in which motions were sufficiently terminated was reached (the time
needed was referred to as $T$). As a termination condition for the process of numerical integration a requirement for the instantaneous
bead velocity (averaged over all beads) to drop below a prescribed threshold was imposed in the simulations. The condition was
$\frac{1}{N}\sum_{i}^{N}|v_{i}|<10^{-6}$.
\newline

During the process of evolutionary optimisation in the two designed networks the improvement of of the allosteric response was recorded.
It is shown in Fig. S1. Details are given in the figure caption.

\section*{Strain propagation and communication pathways}
To discuss the anisotropy of the strain distribution in the random and designed networks we have introduced a method to normalise
the strain of elastic links with respect to the distance from the allosteric site. Therefore, we introduced the shortest {\it graph distance}
between a network bead with index $i$ and a bead of the allosteric pocket with index $p$ as the minimal number of links of a path
connecting the two beads, denoted by $D_{ip}$. The shortest {\it graph distance} of an elastic spring connecting beads $i$ and $j$
to the allosteric pocket was then defined as $D_{ij}:=\text{min}\{D_{ip1},D_{ip2},D_{jp1},D_{jp2}\}$, where $p1$ and $p2$ were the
indices of beads of the allosteric pocket and $p3$ and $p4$ were those of the regulated pocket. Next we defined shells $S_{n}$ of links,
each of which contained all those elastic links $(ij)$
that had the same shortest {\it graph distance} $n$ ($n=1,2,\dots$) to the allosteric pocket, i.e. $S_{n}:=\{\text{all links (ij) for which $D_{ij}=n$}\}$.
Hence, each elastic network link $(ij)$ was uniquely assigned to one shell $S_{n}$ and its elastic strain at time $t$ (defined in the Methods
section of the main text) was denoted by $s_{ij}(t,n)$.
After this procedure, obviously all elastic links had been sorted into their respective shells; all links with shortest {\it graph distance} $1$
were in shell $S_{1}$, those with shortest {\it graph distance} $2$ were in shell $S_{2}$, etc.
Now, during a first simulation of the allosteric operation of a designed network - following ligand binding to the allosteric site until the steady
conformation was reached - for each shell $S_{n}$ a maximum absolute strain value $m_{n}=\text{max}\{|s_{ij}(t,n)|\}$ of all links belonging
to the same shell $S_{n}$ was determined and stored. Then, in a repeated simulation of the same network, the strain of each elastic link $(ij)$
was normalised by dividing its value by the maximum absolute strain value $m_{n}$ of the shell $n$, the particular link belonged to,
i.e. $\tilde{s}_{ij}(t)=s_{ij}(t,n)/m_{n}$. The normalised strain $\tilde{s}_{ij}(t)$ was used to visualise the strain propagation
in both designed networks in a time-resolved fashion, presented in Supplementary Movies V2 and V3. Corresponding snapshots are
shown in the main text Fig. 4. In the same simulation we have memorised for each link $(ij)$ the maximum absolute value $m_{ij}$
of its normalised strain, i.e. $m_{ij}=\text{max}{|\tilde{s}_{ij}(t)|}$. Those values were employed to determine the communication
pathways (shown in the main text Fig. 5), which were constructed by considering only those links that were significantly involved
in the strain propagation, imposed by the condition $m_{ij}>\tilde{s}_{t}$, where $\tilde{s}_{t}$ was a prescribed threshold value for the
normalised strain (with its values given in the main text Methods section).\\
In the random elastic networks a similar procedure of strain normalisation was undertaken. However, to compare the propagation 
of strain to that in the two designed networks, the link strain was normalised with respect to the  coefficients $m_{n}$ from the designed network
with symmetric allosteric coupling.\\
For the selected mutant networks the procedure of strain normalisation was also applied to visualise strain propagation during the
allosteric transition (Movies V4-V7) and to construct communication pathways (Fig. 6 main text and Fig. S3).

It should be noted that for the normalisation procedure the shortest {\it graph distance} between a network bead and the a bead of the
allosteric pocket was determined with a standard algorithm using the powers of the adjacency matrix.

\section*{Robustness of allosteric communication}
Robustness of allosteric communication with respect to exemplary single structural mutations applied to each of the two designed
prototype networks was analysed. To determine the communication chains for the two mutants of the designed network with symmetric
allosteric communication shown in Fig. S3, a threshold of $\tilde{s}_{t}=55\%$ was used for the $52-197$ mutant, and $\tilde{s}_{t}=60\%$
was used for the $196-197$ mutant.

\section*{Supporting Movies and Data}

Time-dependent propagation of strain in the random network, the two designed networks, and the selected mutant networks are provided
as Supporting Movies V1-V7.
In each of the movies the frame rate is not constant and has been adapted such the fast dynamics inside the allosteric pocket, the
inter-domain propagation, as well as the slow motions inside the regulated pocket, can be conveniently followed. The actual time
during the simulation is always given at the bottom right in the movies.

\begin{itemize}
\item {\bf Supporting Movie V1} Conformational motions and strain propagation in the allosterically inactive initial random elastic network.
\item {\bf Supporting Movie V2} Conformational motions and strain propagation in the designed network with symmetric allosteric coupling.
\item {\bf Supporting Movie V3} Conformational motions and strain propagation in the designed network with asymmetric allosteric coupling.
\item {\bf Supporting Movie V4} Conformational motions and strain propagation in the $52-197$ mutant of the designed network with symmetric
allosteric coupling.
\item {\bf Supporting Movie V5} Conformational motions and strain propagation in the $196-197$ mutant of the designed network with symmetric
allosteric coupling.
\item {\bf Supporting Movie V6} Conformational motions and strain propagation in the $52-197-196$ double-mutant of the designed network with
symmetric allosteric coupling.
\item {\bf Supporting Movie V7} Conformational motions and strain propagation in the $52-185$ mutant of the designed network with asymmetric
allosteric coupling.
\end{itemize}

As Supplementary Data the set of spatial coordinates $\vec{R}_{i}^{(0)}$ of the initial random network and the two designed networks are
provided in the respective equilibrium conformation.

\begin{itemize}
\item random\_network.dat: text file containing spatial coordinates of the initial random network.
1st column: bead index ($0$ to $199$); 2nd,3rd,4th columns: x,y,z coordinate of the bead position. 
\item designed\_network1.dat: text file containing spatial coordinates of the designed network with symmetric allosteric coupling.
1st column: bead index ($0$ to $199$); 2nd,3rd,4th columns: x,y,z coordinate of the bead position. 
\item designed\_network2.dat: text file containing spatial coordinates of the designed network with asymmetric allosteric coupling.
1st column: bead index ($0$ to $199$); 2nd,3rd,4th columns: x,y,z coordinate of the bead position. 
\end{itemize}

\begin{figure*}[h!]
\centering{\includegraphics[width=12cm]{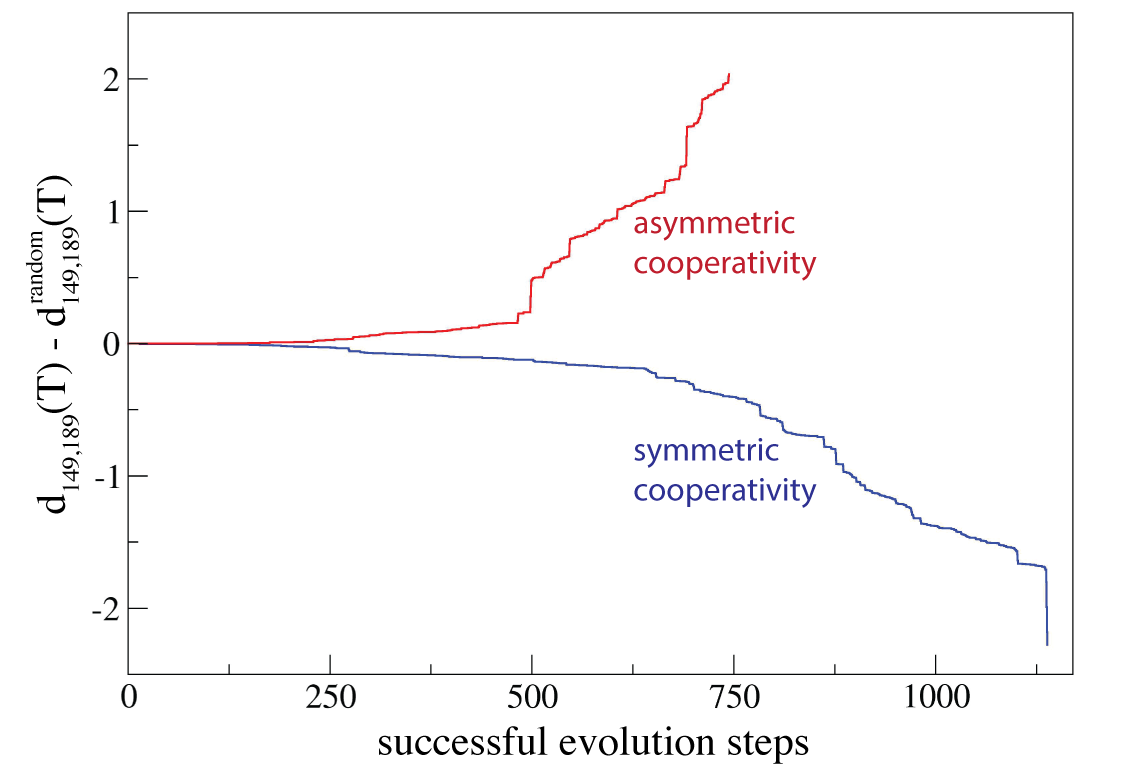}}
\caption{Improvement of the allosteric response in the evolving networks during the two independent processes of evolutionary optimisation;
$d_{149,189}(T)$ is the allosteric parameter of the elastic network at the current stage of evolution and $d^{random}_{149,189}(T)$ is the
value in the initial random elastic network.  A successful evolution step corresponded to an accepted mutation, i.e. one after which the allosteric response
of the network had improved. The total number of successful evolution steps was $1,138$ in the design of the network with symmetric
cooperativity and $744$ in the design of the network with asymmetric cooperativity, respectively.}
\end{figure*}

\begin{figure*}[h!]
\centering{\includegraphics[width=14cm]{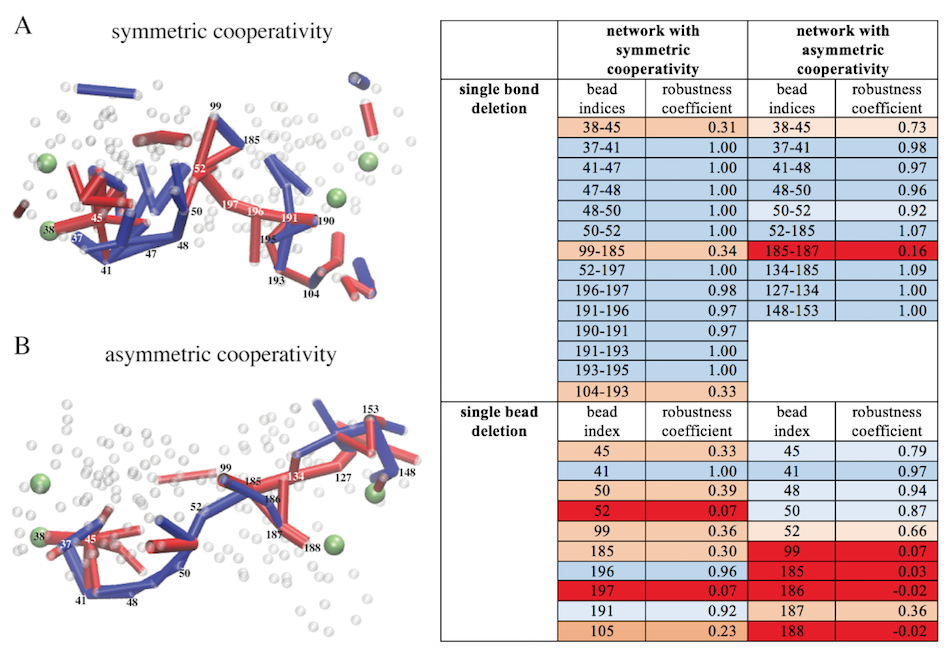}}
\caption{and Table T1. {\bf Mutations and robustness of allosteric communication.} A,B): Communication chains in the designed {\it wildtype} networks
are shown in the same perspective as in Fig. 5 (main text), with the beads which were subjected to mutations being indicated by the
respective bead indices. Mutations performed in both designed networks are listed in the table on the right side. A colour scale from blue
to red shall illustrate neutral to fatal mutations, according to the respective robustness coefficient (defined in the main text Methods section).
For the $52-197-196$ double mutant of the network with symmetric coupling (which is not listed in the Table) the robustness coefficient was $0.99$.}
\end{figure*}

\begin{figure*}[h!]
\centering{\includegraphics[width=14cm]{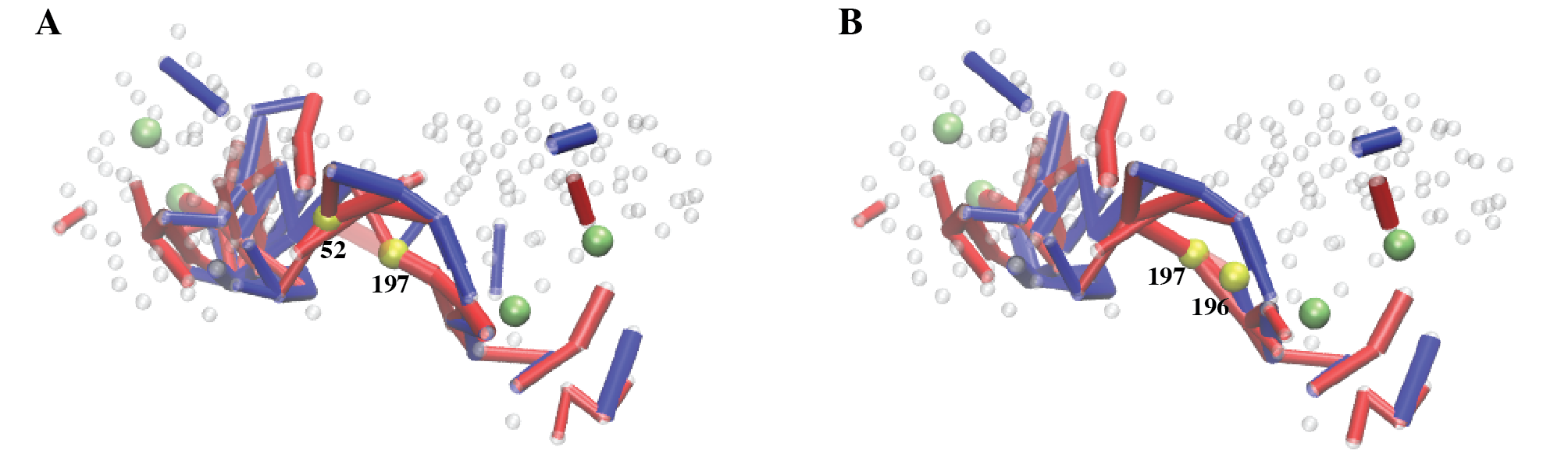}}
\caption{{\bf Communication chains in selected mutant structures.} For the designed network with symmetric cooperativity the communication
chains in two representative mutants, each in which a single elastic link was deleted, are shown. The $52-197$ mutant (A) as well as the
$196-197$ mutant (B) both retained full allosteric activity, with robustness coefficients of $1.00$ and $0.98$, respectively (see Table T1). In
each mutant the deleted link is shown in transparent and the corresponding beads are highlighted in yellow.}
\end{figure*}

\end{document}